\documentclass[3p,twocolumn,times]{elsarticle}

\usepackage{ecrc}

\volume{00}

\firstpage{1}

\journalname{Nuclear Physics B Proceedings Supplement}

\runauth{}

\jid{nuphbp}

\jnltitlelogo{Nuclear Physics B Proceedings Supplement}

\usepackage{amssymb}
\usepackage{amsmath}

\usepackage[figuresright]{rotating}

\begin{document}

\begin{frontmatter}



\dochead{}

\title{Electrically Millicharged Neutrino in Media}


\author[address_1]{I. Balantsev}
\ead{balantsev@physics.msu.ru}

\author[address_1,address_2]{A. Studenikin}
\ead{studenik@srd.sinp.msu.ru}

\address[address_1]{Department of Theoretical Physics, Moscow State University, Moscow, Russia}
\address[address_2]{Joint Institute for Nuclear Research, Dubna, Russia}

\begin{abstract}
On the basis of nonzero neutrino electromagnetic properties, we consider a problem of electrically millicharged neutrino energy
spectra in a magnetized matter. It is shown that in this case neutrino energies are quantized. These phenomena can be important for astrophysical
applications.
\end{abstract}


\end{frontmatter}


Within the Standard Model a massless neutrino has trivial (or vanishing) electromagnetic properties. However, a massive neutrino \footnote{There is
no doubt that the recent studies of flavour conversion in solar, atmospheric, reactor and accelerator neutrino fluxes give strong evidence for
non-zero neutrino mass.} even in the easiest generalization of the Standard Model should have non-trivial electromagnetic properties
\cite{MarSanPLB77LeeShrPRD77FujShrPRL80}. For a recent review on neutrino electromagnetic properties see \cite{GiuStuPAN09}. A massive neutrino can
also be electrically charged (millicharged) particle in a class of more general theoretical models \cite{FooLewVolJPG93_BabMohPRD89}. The most severe
experimental constraints on the electric charge of the neutrino, $q_\nu \leq 10^{-21} e$ \cite{MarMorPLB84}.

We consider a millicharged massive neutrino propagating in nonmoving magnetized medium composed of neutrons using the modified Dirac equation for the
neutrino wave function exactly accounting for the neutrino interaction with matter \cite{StuTerPLB05GriStuTerPLB05GriStuTerG&C05GriStuTerPAN06} where
we should substitute the neutrino momentum $p^{\mu}$ for the "extended" momentum $p^{\mu}\rightarrow p^{\mu}-q_{\nu}A^{\mu}$ to account for the
neutrino interaction with magnetic field:
\begin{equation}\Big\{
\gamma_{\mu}(p^{\mu}-q_{\nu}A^{\mu})+\frac{1}{2} \gamma_{\mu}(1+\gamma^5)f^{\mu}-m \Big\}\Psi(x)=0,\nonumber
\end{equation}
where $q_{\nu}$ is a millicharge of the neutrino. Here we choose that the electromagnetic field and effective matter potential are
$A^{\mu}=(0,-\frac{yB}{2},\frac{xB}{2},0),\,f^{\mu}=-Gn(1,0,0,0),$ where $G=\frac{G_F}{\sqrt{2}}$, $n$ is matter number density. Than we can get the
neutrino energy spectrum,
\begin{equation}\hspace{0.5cm}
p_0=\frac{Gn}{2}+\varepsilon\sqrt{\left(-\frac{Gn}{2}+mT^0\right)^2+m^2},\quad\varepsilon=\pm 1,\nonumber
\end{equation}
where $T^0=\frac{s'}{m}\sqrt{p_3^2+2q_{\nu}BN},s'=\pm 1$ are eigenvalues of the spin operator
$\hat{T}^0=\frac{1}{m}\boldsymbol{\sigma}(\hat{\bold{p}}-q_{\nu}\bold{A})$ that commutes with the corresponding Hamiltonian, $N=0,1,2,...$

From this spectrum it follows that the effect of the neutrino
trapping on circular orbits in magnetized matter exist that can be
important for astrophysical applications.

\textbf{Acknowledgements.} One of the authors (A.S.) is thankful to Professor George Tzanakos for the kind invitation to participate in the Neutrino 2010 conference.


\section*{References}



\bibliographystyle{elsarticle-num}
\bibliography{}



\end{document}